\input{psfig.sty}
\documentclass{aa}

\begin{document}

   \thesaurus{06     
              (03.11.1;  
               16.06.1;  
               19.06.1;  
               19.37.1;  
               19.53.1;  
               19.63.1)} 
\title{On the Spectral Evolution of Cygnus~X--2 along its Color-Color Diagram}

  \subtitle{}

   \author{T. Di Salvo \inst{1} \and
   R. Farinelli\inst{2} \and L. Burderi\inst{3} \and F. Frontera\inst{2,4} \and
   E. Kuulkers\inst{5,6} \and N. Masetti\inst{4} \and N.R. Robba\inst{7} \and
   L. Stella\inst{3} \and M. van der Klis\inst{1}
          }

   \offprints{T. Di Salvo}

   \institute{Astronomical Institute "Anton Pannekoek," University of 
	      Amsterdam and Center for High-Energy Astrophysics,
	      Kruislaan 403, NL 1098 SJ Amsterdam, the Netherlands\\
              email: disalvo@astro.uva.nl
       \and
           Physics Department, University of Ferrara, Via
	     Paradiso, 12, 44100 Ferrara, Italy
	 \and
	     Osservatorio Astronomico di Roma, Via Frascati 33, 
	     00040 Monteporzio Catone (Roma), Italy    
	 \and
	     Istituto TESRE, CNR, Via P. Gobetti, 101, 40129 
	     Bologna, Italy    
	 \and 
	     SRON National Institute for Space Research, Sorbonnelaan 2, 
	     3584 CA Utrecht, the Netherlands    
	 \and
	     Astronomical Institute, Utrecht University, P.O. Box 80000, 
	     3508 TA Utrecht, the Netherlands    
	 \and 
	     Dipartimento di Scienze Fisiche ed Astronomiche, 
	     Universit\`a di Palermo, via Archirafi 36 - 90123 Palermo, Italy    
             }

   \date{Received 21/11/01; accepted 12/02/02}

   \maketitle

\begin{abstract}

We report on the results of a broad band (0.1--200 keV) spectral
study of Cyg X--2 using two BeppoSAX observations taken in 1996 and 
1997, respectively, for a total effective on-source time of $\sim 100$ ks.
The color-color (CD) and hardness-intensity (HID) diagrams 
show that the source was in the horizontal branch (HB) and normal branch 
(NB) during the 1996 and 1997 observation, respectively. 
Five spectra were selected around different positions 
of the source in the CD/HID, two in the HB and three in the NB.
These spectra are fit to a model consisting of a disk blackbody, a 
Comptonization component, and two Gaussian emission lines at $\sim 1$ keV
and $\sim 6.6$ keV, respectively. The addition of a hard power-law
tail with photon index $\sim 2$, contributing $\sim 1.5\%$ of the source 
luminosity, improves the fit of the spectra in the HB.
We interpret the soft component as the emission from the inner accretion
disk, with inner temperature, $k T_{\rm in}$, varying between $\sim 0.8$ and
$\sim 1.7$ keV and inner radius, $R_{\rm in}$, varying between $\sim 26$ and
$\sim 11$ km (assuming an inclination angle of the system of $60^\circ$).
The Comptonization component is probably emitted by hot plasma (electron
temperature $k T_{\rm e}$ varying between $\sim 3$ and $\sim 20$ keV, optical
depth $\tau \sim 11-0.4$, seed-photon temperature $kT_{\rm W} \sim 1-2.4$ keV)
surrounding the NS. The changes in the parameters
of the blackbody component indicate that the inner rim of the disk
approaches the NS surface when the source moves from the HB to the NB, i.e.\
as the (inferred) mass accretion rate increases.
The parameters of the Comptonized component also change significantly when
the source moves from the HB to the NB. 
We discuss possible scenarios which can explain these changes.


\end{abstract}

\section{Introduction}

Low Mass X-ray Binaries (hereafter LMXBs) are binary systems where a low-mass
star loses mass which is accreted, at least in part, onto a weakly magnetic 
neutron star (NS) or a black hole candidate (BHC). 
LMXBs containing NSs are usually divided into two classes:
the so-called Z sources, with luminosities close to the Eddington luminosity, 
$L_{\rm edd}$, and the Atoll sources, which usually have lower luminosities, 
$\sim 0.01-0.1\ L_{\rm edd}$.  This widely used classification relies upon
a combination of the X-ray spectral properties of these sources, namely the
pattern traced out by individual sources in an X-ray color-color diagram
(CD) or hardness-intensity diagram (HID), and their correlated timing
properties (Hasinger \& van der Klis 1989). The six known (Galactic) Z
sources describe a Z-track in the CD on timescales of a few days, while 
the sources of the atoll class trace out an atoll-shaped track in the CD 
on a timescale of weeks. Considerable evidence has been found that the 
mass accretion rate of individual Z-sources increases from the top left to 
the bottom right of the Z-pattern ({\it e.g.} Hasinger et al. 1990), 
i.e. along the so called horizontal, normal and flaring branches 
(hereafter HB, NB and FB, respectively). Similarly, in atoll sources 
the accretion rate increases from the so-called island state
to the upper banana branch. 

Accurate timing studies, mainly performed with the large area Proportional 
Counter Array (PCA) on board the RXTE satellite, have shown the existence of 
compelling correlations between rapid variability phenomena -- such as 
Quasi-Periodic Oscillations, QPOs, band-limited noise, etc., present in the 
frequency range extending from Hz to kHz -- and spectral states as defined by 
the position of the source on the CD (cf. Hasinger \& van der Klis 1989; for 
reviews see van der Klis 1995, 2000). 
In fact (almost) all the timing features characterizing the power density
spectra of these sources seem to vary in a  
smooth and monotonic way when the source moves along its CD track
(e.g. Wijnands \& van der Klis 1999; Psaltis et al.
1999); this clearly indicates a tight correlation between 
spectral and temporal behavior. 
These results have strengthened the idea that a single basic parameter
determines both the temporal and the spectral behavior of these
sources, and that this is most probably the mass accretion rate.
However, the lack of a direct correlation between X-ray luminosity 
(where most of the accretion energy should be released) and the {\it inferred} 
mass accretion rate (e.g. Hasinger \& van der Klis 1989; M\'endez 2000, 
and references therein), is still to be understood. Indeed the observed 
X-ray flux may be easily affected by geometrical effects. In this case 
multiwavelength observations, from X-rays to UV and optical (which are thought 
to originate from X-rays reprocessed by the accretion disk, which subtends a 
large solid angle as seen from the X-ray source and is therefore less affected 
by geometrical effects), are required to properly study the influx of X-rays 
and therefore the accretion rate onto the compact object.  Multiwavelength
observations conducted on some Z sources have already shown a good correlation
between X-ray spectral states and UV emission, indicating that the mass 
accretion rate increases from the HB to the NB and to the FB (see e.g.\ 
Hasinger et al. 1990; Vrtilek et al. 1990; 1991a).  Alternatively,    
the mass accretion rate might vary non-monotonically along the CD track, a 
possibility suggested by some recent results on the decrease of the frequency 
of the so-called horizontal branch oscillations, HBOs, in 
GX 17+2 when the source was in the NB, i.e.\ at high inferred mass accretion 
rates, while simultaneously the frequency of the kHz QPOs, which are thought 
to track the inner radius of the disk, continued increasing 
(Homan et al. 2002). Also, in some cases, the X-ray burst properties, which
are expected to depend on the mass accretion rate, do not show a clear 
correlation with the position in the CD (see e.g.\ Kuulkers et al. 2002). 
A possible explanation for
this phenomenology, where the X-ray flux can directly reflect accretion rate,
$\dot M$, and the spectral and temporal properties depend on $\dot M$
normalized by its own long-term average, which determines the inner radius of 
the disk, was recently proposed by van der Klis (2001).
 
The X-ray spectra (and also timing properties) of the atoll sources show
some similarity with those of BHCs in their lower luminosity states. 
In fact these spectra can often be described by a soft component, 
fitted by a blackbody or a multi-temperature disk blackbody
($kT \sim 0.5-1$ keV), plus a power law with high energy cutoff usually at
energies between $\sim 10$ and $\sim 100$ keV, probably originating from
thermal Comptonization of soft photons by hot electrons in a surrounding 
corona (see, however, Markoff et al. 2001). 
Also, atoll sources can be found in soft (high luminosity) spectral states, 
where most of the energy is emitted below $\sim 10$ keV, and hard (low 
luminosity) spectral states, where the soft emission is much 
reduced and the spectrum is dominated at high energies by a power 
law.  It was originally thought that the electron temperature in the 
scattering cloud should be lower for the NSs than for the BHCs
because of the extra cooling due to the soft photons emitted by the NS 
surface (e.g.\ Churazov et al. 1997). 
This seems indeed true for those systems in which a high energy
cutoff has been observed, although some LMXBs harbouring a NS 
do not show any evidence for a cutoff up to energies of $\ge 100$~keV 
(Barret et al. 1991; Yoshida et al. 1993; Harmon et al. 1996; 
Piraino et al. 1999; Barret et al. 2000).

The high energy spectral hardness of LMXBs, i.e.\ the ratio of the count
rate in the 40--80~keV band to that in the 13--25~keV band, decreases
monotonically for sources of increasing luminosity (van Paradijs \&
van der Klis 1994). 
In fact the X-ray spectra of the Z sources are softer than those of
atoll sources and dominated by thermal-like components with
temperatures well below $\sim 10$~keV. 
This is in agreement with the expectation that in a high-luminosity 
regime the presence of numerous soft photons could provide strong Compton 
cooling and result in softer spectra.
However, hard power-law components, dominating the spectrum above 
$\sim 30$ keV, have been detected in several luminous and otherwise
soft Z sources. These were detected several times in the past, e.g.\ 
in Sco X--1 (e.g.\ Peterson \& Jacobson 1966), 
and in GX 5--1, where the power-law intensity decreased when the source
moved from the NB to the FB, i.e., from lower to higher inferred mass 
accretion rates (Asai et al. 1994; note that contamination from a nearby
source could not be excluded in this case).  Hard power-law tails have 
been recently found in several other bright LMXBs.
These components can be fit by a power law, with photon index in the 
range 1.9--3.3, contributing from 1\% to 10\% of the source luminosity;  
significant variations with the spectral state of the source have been
detected. 
In particular, the first unambiguous detection of a hard tail varying with
the position of the source in the CD was obtained through 
a 0.1--200~keV BeppoSAX observation of GX~17+2 
(Di Salvo et al. 2000), where a hard power-law 
component was observed in the HB, which systematically weakened 
(by up to a factor of $\sim 20$) when the source moved to the NB. 
The presence of a variable hard tail in Sco X--1 was confirmed by OSSE and 
RXTE observations (Strickman \& Barret 2000; D'Amico et al. 2001). 
A hard tail was also detected in BeppoSAX data of GX~349+2 (Di Salvo et al. 
2001), Cyg~X-2 (Frontera et al. 1998), and the peculiar 
LMXB Cir X--1 (Iaria et al. 2001). 
In most of the cases cited above the hard component appeared to become weaker 
at higher inferred accretion rates.  The only known exception to this
behavior is currently Sco X--1, where the presence and intensity of the
hard power-law tail does not show any clear correlation 
with the position of the source in the CD (D'Amico et al. 2001); it
might, however, be correlated with periods of radio flaring
(Strickman \& Barret 2000).

Among the Z sources Cyg X--2 is one of the most interesting and 
best-studied sources.  The distance and the inclination angle of the
source have been estimated to be 8 kpc and $60^\circ$, respectively
(see e.g.\ Hjellming et al. 1990; Orosz \& Kuulkers 1999).  
The companion is an evolved, late-type $\sim 0.4-0.7\;M_\odot$ star, the 
spectral type of which seems to vary from A5 to F2 with the binary period 
of 9.84 days, probably due to the heating of the companion surface by the 
X-ray emission from the compact object (see, however, Casares et al. 1998).
The latter was identified as a low magnetic field NS after the observation 
of type-I X-ray bursts from the system (e.g.\ Kahn \& Grindlay 1984).
Cyg X--2 shows high-intensity states (with a 2--10 keV luminosity
higher than $7 \times 10^{37}$ ergs/s), usually characterized by 
irregular dipping activity (Vrtilek et al. 1988; see also Kuulkers 
et al. 1996). This dipping activity is absent during the low-intensity 
states, which instead show a modulation of the X-ray flux, i.e.\ an intensity 
reduction up to 40\% at the binary phase 0.4--0.8, probably due to the 
obscuration of an accretion disk corona by a tilted accretion disk 
(Vrtilek et al. 1988).
Cyg X--2 usually exhibits all three branches of the Z-track,
although its behavior in the FB is quite unusual. It shows
an intensity decrease from the bottom of the NB to the FB which can
be identified with the dipping activity.
Cyg X--2, together with GX 5--1 and GX 340+0, shows secular variations 
of the position of the Z-pattern in the CD on long time scales 
(more than a few days; see Kuulkers et al. 1996 for an extensive study).
These variations have been interpreted in terms of a high inclination angle
of these sources, allowing matter near the equatorial plane to 
modify the emission from the central region (e.g.\ Kuulkers et al. 1996), 
but these could also be a consequence of a more general behavior of LMXB 
accretion (van der Klis 2001).
Simultaneous multiwavelength observations showed that the UV flux
and optical emission lines' strength increase from the HB to the FB (Vrtilek 
et al. 1990; van Paradijs et al. 1990).  Also, strong non-thermal
radio flares were observed in the HB and upper NB, whereas the source
was radio quiet in the lower NB and in the FB (Hasinger et al. 1990).

The continuum spectrum of Cyg X--2 has been fitted by two-component
models, in particular the so-called Western model (White et al. 1986), 
consisting of a blackbody component, originating from (or close to) the NS, 
and a power law with high energy cutoff, approximating unsaturated 
Comptonization from the inner disk (see e.g.\ Hasinger et al. 1990 using Ginga
data; Smale et al. 1993 using BBXRT data), and the Eastern model 
(Mitsuda et al. 1984), consisting of a multi-temperature blackbody from the 
accretion disk and a Comptonized blackbody from the NS (e.g.\ Hasinger et al. 
1990; Hoshi \& Mitsuda 1991 using Ginga data). 
In the framework of the Eastern model, the disk emission has also been 
described by non-standard models, where the radial temperature gradient $q$
was considered as a free parameter in the fit  
(Hirano et al. 1995). This parameter was found to change continuously 
from the HB to the FB, suggesting that the disk undergoes structural
changes when the source moves along the CD.
However, the evolution of the spectral parameters of the source along the 
CD track is still not clear, given that the data collected and examined up 
to now did not allow distinguishing between the different models, which 
in turn led to different results.
The Cyg X--2 X-ray spectrum above $\sim 20$ keV has been studied with
detectors on balloons (an overview of the earliest high energy 
observations of Cyg X-2 can be found in Peterson 1973).
Interestingly, an unexpectedly hard spectrum was observed from Cyg X--2.
This was fitted by a power law with photon index 2.8 (Maurer et al.
1982) or 1.9 (Ling et al. 1996).  Matt et al. (1990) concluded that,
while the 25--50 keV flux from Cyg X--2 was compatible with the extrapolation
of the low energy spectrum, the shape of the hard spectrum was flatter than
expected, and that therefore only simultaneous broad band observations of
the X-ray spectrum could confirm or disprove the presence of a third hard
component. 

An emission line at $\sim 6.7$ keV from highly ionized iron has also 
been reported in the energy spectrum. 
The line is quite broad (its width was measured from BBXRT data to be 
$\sim 1$ keV FWHM) and has an equivalent width of $\sim 60$ eV
(Smale et al. 1993). These authors suggest that the width of the line, 
together with the lack of a significant iron absorption edge, could be
explained by reflection from a highly ionized disk 
surface. An Fe absorption edge, with an energy of 8 keV and column 
density of $\sim 10^{19}$ cm$^{-2}$, was reported 
in the lower NB and in the FB from Ginga observations (Hirano et al. 1995).
Note, however, that these features are usually quite broad and therefore 
sensitive to the assumed continuum model.
Emission lines at energies between $0.6-1.1$ keV were reported several times  
(see e.g.\ Vrtilek et al. 1986, 1991b; Kuulkers et al. 1997); these likely 
result from ionized O and Fe.
The strength and energy of these soft lines appears to depend on the spectral 
state of the source.  This is to be expected given that these lines are very 
sensitive to the temperature of the emitting material (Kallman et al. 1989).  

We report here on the broad band (0.1--200 keV) spectral analysis of two 
BeppoSAX observations of Cyg X--2. The source was in the HB and NB during 
these observations, providing the possibility of studying the spectral 
variations along a considerable part of the CD.  
A hard power-law tail is significantly detected only in the HB spectra, 
showing a behavior similar to that observed in GX 17+2.

\section{Observations}

We analyzed two BeppoSAX (Boella et al. 1997a) observations of Cyg X--2 from 
the public BeppoSAX archive. The first observation was performed during the 
Science Verification Phase on 1996 July 23 for an exposure time of 
$\sim 33$ ks.  The second observation took place in 1997 October 
26--28, for an exposure time of $\sim 66$ ks.
Results from the 1996 and 1997 observations were discussed, respectively, 
in Kuulkers et al. (1997, who analyzed the 0.1--10 keV spectrum from the 
low-energy instrument) and in Frontera et al. (1998, who analyzed the 
2--200 keV spectrum, averaged over the whole observation).  
The BeppoSAX Narrow Field Instruments (NFI) consist of four co-aligned 
instruments covering the energy range between 0.1 and 200 keV.  
X-ray images were obtained with three (in July 1996) and two (in October
1997) Medium Energy Concentrator Spectrometers (MECS, position sensitive
gas scintillator proportional counters operating in the 1.3--10.5~keV band; 
Boella et al. 1997b; in May 1997 detector unit 1 stopped working) and a 
Low Energy Concentrator Spectrometer (LECS, a thin-window position-sensitive 
gas scintillator proportional counter with extended low energy response, 
0.1--10~keV; Parmar et al. 1997), each in the focal plane of the X-ray 
concentrators. 
These in turn consist of a set of 30 nested conical mirror shells providing 
an approximation to a Wolter-I type geometry.  
Due to UV contamination problems, the LECS was operated only at satellite 
night time, resulting in reduced exposure time ($\sim 10.2$ and $\sim 23$ ks 
for the 1996 and 1997 observations, respectively). The LECS and MECS 
Cyg X--2 data were extracted from circular regions centered on the source 
and with radii of $8'$ and $4'$, respectively,
corresponding to $\sim 90$\% of the instruments' point spread function. 
Data extracted from the same detector regions during blank field 
observations were used for background subtraction.
The High Pressure Gas Scintillation Proportional Counter 
(HPGSPC; energy range of 6--30 keV; Manzo et al. 1997) and
Phoswich Detection System (PDS; energy range of 13--200 keV; Frontera et al.
1997) are non-imaging instruments, the field of view ($\sim 1^\circ$ FWHM) 
of which is limited by rocking collimators that can alternate on and off the 
source.  Background subtraction for these instruments was obtained from 
data accumulated during off-source intervals. 
The effective exposure was 20 and 26~ks in the HPGSPC and 18 and 27~ks
in the PDS during the 1996 and 1997 observations, respectively.
No contaminating sources are known to be present within $1^\circ$ from
Cyg X--2, and
all the available evidences confirm that there were no significant
contaminating sources in the field of view of the HPGSPC and PDS, both in the 
on-source and off-source positions.  In fact the off-source count rates 
are in the expected range and display the characteristic orbital dependence 
arising from the variable particle background; moreover 
the HPGSPC and PDS spectra align well with each other and with the MECS
spectra. 

\begin{figure*}
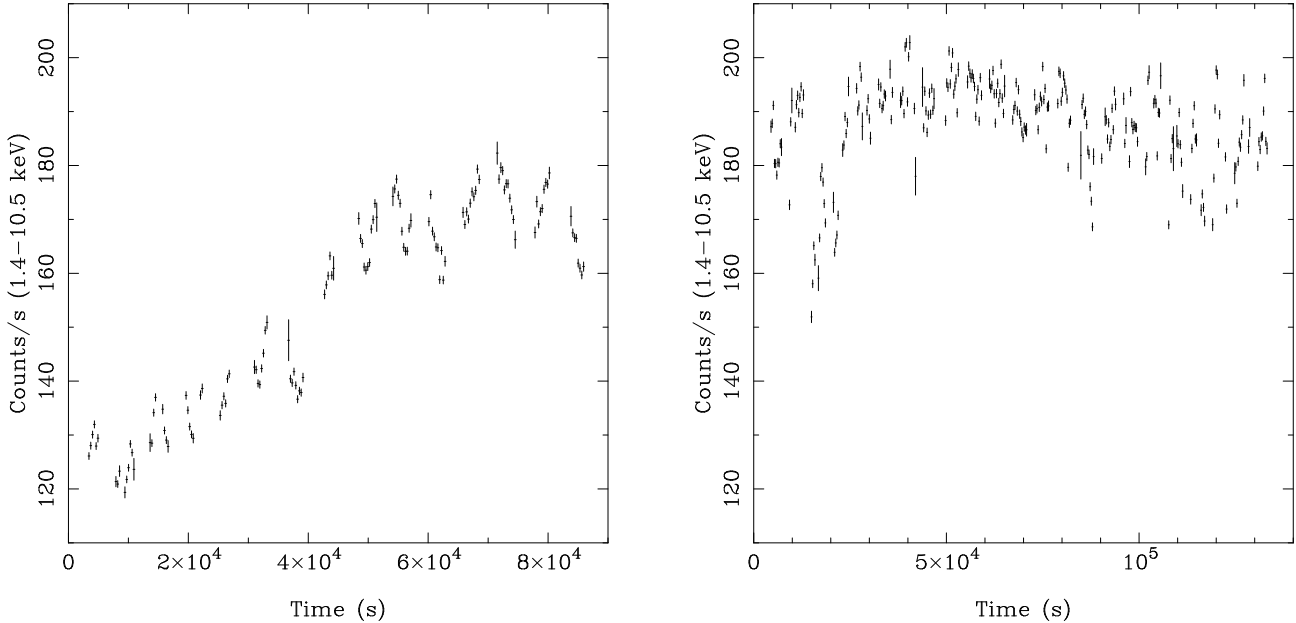

\vspace{0cm}
\hbox{\hspace{0cm}\psfig{figure=fig1a.ps,width=8.0cm}\hspace{1cm}
\psfig{figure=fig1b.ps,width=8.0cm}}
\vspace{0cm}
\hfill      \parbox[b]{18cm}{\caption[]{Light curve of Cyg X--2 in the 
1.4--10.5~keV (MECS data) during the 1996 observation (starting time July 
23 00:54:21 UT, left panel) and the 1997 observation (starting time 
October 26 15:37:11 UT, right panel), respectively.
Each bin corresponds to 300 s integration time.
}\label{fig1}}%
\end{figure*}

\begin{figure*}
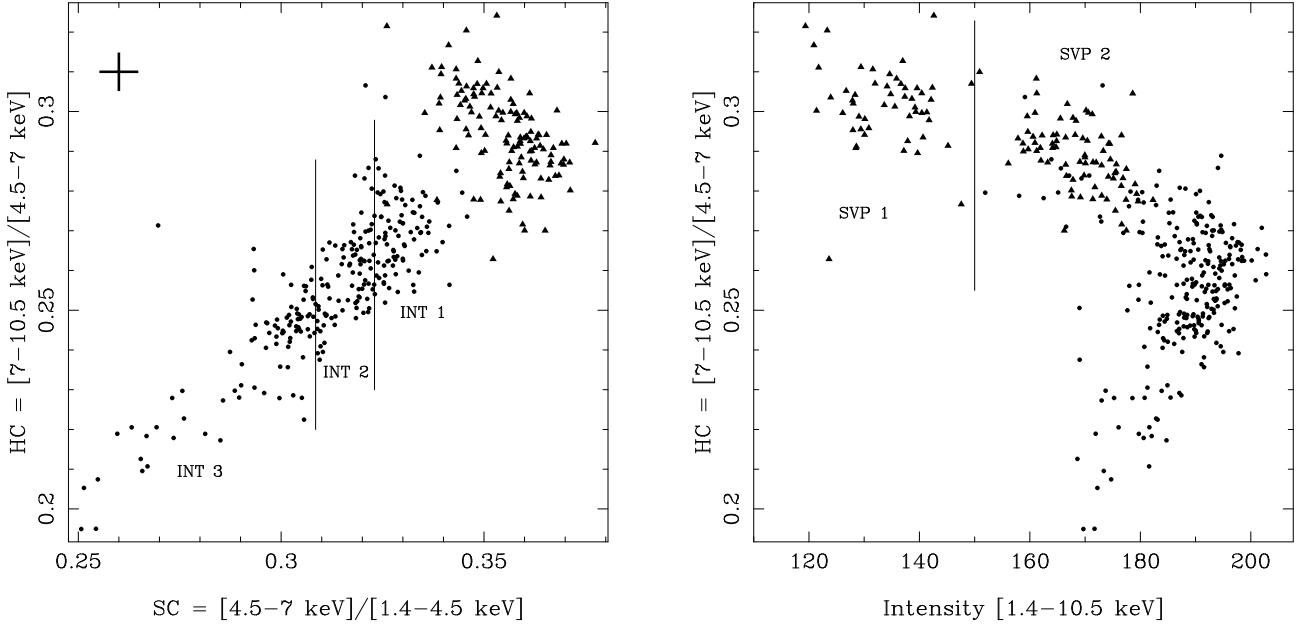

\vspace{0cm}
\hbox{\hspace{0cm}\psfig{figure=fig2a.ps,width=8.0cm}\hspace{1cm}
\psfig{figure=fig2b.ps,width=8.0cm}}
\vspace{0cm}
\hfill      \parbox[b]{18cm}{\caption[]{Color-color diagram (left) and 
hardness-intensity diagram (right) of Cyg X--2 during the 1996 observation 
(triangles) and the 1997 observation (circles).
The hard color (HC) is the ratio of the counts in the energy bands 7--10.5 keV 
and 4.5--7 keV, the soft color (SC) is the ratio of the counts in the energy 
bands 4.5--7 keV and 1.4--4.5 keV, the source intensity is the count rate in 
the MECS (energy range 1.4--10.5~keV). 
Each point corresponds to 300 s integration time.
A typical error bar is shown at the top left corner of the left panel.
The intervals in which the CD/HID were divided for the spectral analysis
(see text) are also shown.
}\label{fig2}}%
\end{figure*}

The light curve of Cyg X--2 in the MECS energy range (1.4--10.5 keV) 
during the 1996 and 1997 observation is shown in Fig.~\ref{fig1} 
(left and right panel, respectively).  During the 1996 observation 
the intensity gradually increased from $\sim 120$ to $\sim 180$ counts/s, 
while during the 1997 observation it was quite constant at $\sim 190$ counts/s, 
except for a ``dip'' in the first part of the observation, during which the 
intensity decreased to $\sim 160$ counts/s.  Fig.~\ref{fig2} shows the CD 
(left) and the HID (right) where the hard color, HC, is the ratio of the 7--10.5
keV and 4.5--7 keV MECS count rates, the soft color, SC, is the ratio of the
4.5--7 keV and 1.4--4.5 keV MECS count rates and the intensity is again the
1.4-10.5~keV count rate.  
Two branches are clearly seen in the diagrams, closely resembling the HB and
NB, due to, respectively, the 1996 and 1997 data. The two branches seem
to connect in the usual way, so there are no obvious shifts in the CD
or HID.
Note that the source intensity variations in the HB were relatively
pronounced.  The 1996 observation was therefore divided into two intervals, 
corresponding to different source intensity ranges, as indicated in 
Fig.~\ref{fig2} (right panel). We will refer to 
these two intervals as the upper HB (i.e.\ the interval corresponding to the
lower intensity range in the HB) and the lower HB (the interval corresponding 
to the higher intensity level in the HB).
During the 1997 observation the source was in the NB, and larger changes in 
the colors took place, while the variations in intensity were less pronounced.
This observation was divided into three time intervals, in which the
source, on the average, exhibited different SC, even if on short
time scales the behavior of the SC was more complex. These intervals are 
shown in Fig.~\ref{fig2} (left panel). 
We call these intervals lower NB (corresponding to lower SC values in the NB), 
middle NB (corresponding to intermediate SC values in the NB) and upper NB 
(corresponding to higher SC values in the NB).
In order to investigate the changes of the source spectrum as a function of 
position in the CD or HID, we accumulated the energy spectra of each NFI 
over the five different intervals described above, i.e.\ 
two spectra in the HB and three in the NB. 
The (absorbed) source luminosity in the 0.1--100 keV energy band 
ranged from $\sim 0.89 \times 10^{38}$ ergs/s
in the upper HB to $\sim 1.5 \times 10^{38}$ ergs/s in the NB, assuming
a source distance of 8 kpc.

In our spectral analysis we used data corresponding to the following energy 
ranges: 0.12--4 keV for the LECS, 1.8--10 keV for the MECS, 8--30 keV for the 
HPGSPC and 15--200 keV for the PDS. All spectra were rebinned in order to 
oversample the full width at half maximum of the energy resolution by a 
factor of 5.  
A systematic 1\% error was added to each spectrum to account for calibration
uncertainties. 
As customary, in the spectral fitting procedure we allowed for different
normalizations in the LECS, HPGSPC and PDS spectra relative to the MECS
spectra, and checked  {\it a posteriori} that the derived values are in the
standard range for each instrument.\footnote{see the BeppoSAX handbook at
ftp://sax.sdc.asi.it/pub/sax/doc/software\_docs.}

\section{Spectral analysis}

In line with previous studies (e.g.\ Mitsuda et al. 1984; White et al. 1988;
Di Salvo et al. 2000, 2001), we tried several two-component models to fit 
the X-ray continuum of Cyg X--2.
These were a blackbody or a multi-temperature blackbody disk model
({\tt diskbb} in the terminology of the spectral fitting program XSPEC v.11) 
for the soft component, together with a blackbody or a thermal 
Comptonization model for the harder component.  Among Comptonization spectra 
we considered:  a power law with exponential cutoff ({\tt cutoffpl}),  
a Comptonization spectrum of cool photons scattering off hot electrons 
({\tt compst}, based on the solution of the Kompaneets equation given by 
Sunyaev \& Titarchuk 1980), and a Comptonization spectrum of an input Wien 
spectrum of soft photons by a hot plasma ({\tt comptt}, which includes 
relativistic effects and takes into account the dependence of the scattering 
opacity on seed-photon energy, which is a free parameter of the model; 
Titarchuk 1994; Hua \& Titarchuk 1995; Titarchuk \& Lyubarskij 1995).

For the three spectra in the NB the best fit to the continuum was found using 
the {\tt diskbb} plus {\tt comptt} model (see results in Table~1). 
This is slightly different from the blackbody plus {\tt comptt} model 
that was used to fit the spectra of other Z sources (e.g.\ Di Salvo et al. 
2000; Di Salvo et al. 2001); in the case of Cyg X--2 this model gave
significantly worse fits for some of the intervals.
In fact by using this model we obtained $\chi^2/d.o.f.$
of 274/190, 398/190 and 273/189 for the spectra of the upper NB, the middle 
NB and the lower NB, respectively, generally worse (especially in the case
of the middle NB spectrum) than the results obtained using {\tt diskbb}
instead of blackbody (cf.\ Table~1).  The fit with the {\tt diskbb} is 
preferable also because using the blackbody for the soft component
the low energy line becomes unphysically broad ($\sigma \sim 0.32$ keV) and
strong (equivalent width of $\sim 230$ eV) and centered at lower energies 
($\sim 0.89$ keV). In the cases of the middle and the lower NB spectra
the fit using the single-temperature blackbody can be improved adding a
power law with photon index $\sim 3$. This model gives $\chi^2/d.o.f.$
of 311/188 and 213/187 for the middle NB and lower NB, respectively, which 
are comparable with the ones obtained using {\tt diskbb} for the soft 
component. Again this last model is preferable because it is simpler and
because the steep power law is indeed needed to fit the soft part of the 
spectrum instead of the hard part, where it gives almost no contribution.
We therefore assumed the {\tt diskbb} plus {\tt comptt} model reported in 
Table~1 as the best fit model.

\begin{table*}
\caption[]{Results of the fit of Cyg X--2 spectra in the energy band 
0.12--200 keV, with a disk blackbody, a Comptonized spectrum modeled by 
{\tt Comptt}, a power law, and two Gaussian emission lines.
Uncertainties are at the 90\% confidence level for a single parameter.
$k T_{\rm W}$ is the temperature 
of the soft seed photons for the Comptonization, $k T_{\rm e}$ is the 
electron temperature, $\tau$ is the optical depth of the 
scattering cloud in a spherical geometry. 
$I$ is the intensity of a Gaussian emission line in units of photons cm$^{-2}$ 
s$^{-1}$.
The power-law normalization is in units of photons keV$^{-1}$ cm$^{-2}$ 
s$^{-1}$ at 1 keV.  Upper limits on the power law normalizations are at 90\%
confidence level.
The total flux, in units of $10^{-8}$ ergs cm$^{-2}$ s$^{-1}$, refers to 
the 0.1--100 keV energy range.
F-test indicates the probability of chance improvement of the fit when the 
power law is included in the spectral model.
}
\begin{center}
\label{tab1}
\begin{tabular}{l|c|c|c|c|c} 
\hline \hline
 Spectrum  & SVP1 &    SVP2     &     INT1   &     INT2    &       INT3   \\
 Position  & UHB  &    LHB      &     UNB    &     MNB     &       LNB    \\
\hline
$N_{\rm H}$ $\rm (\times 10^{22}\;cm^{-2})$ 
& 0.197 (frozen) & $0.197 \pm 0.015$ & $0.1904 \pm 0.0029$ 
& $0.1823 \pm 0.0021$ & $0.1911 \pm 0.0047$ \\

$k T_{\rm in}$ (keV) 
& $0.82^{+0.07}_{-0.15}$ & $0.92 \pm 0.55$ & $1.694 \pm 0.022$ 
& $1.659 \pm 0.019$ & $1.546 \pm 0.030$ \\

$R_{\rm in} \sqrt{\cos i}$ (km)
& $18.6 \pm 3.5$ & $16.1 \pm 2.1$ & $8.08 \pm 0.16$ 
& $8.7 \pm 0.16$ & $9.74 \pm 0.30$ \\

$k T_{\rm W}$ (keV)
& $1.062 \pm 0.12$ & $1.12 \pm 0.12$ & $2.39 \pm 0.39$ 
& $2.285 \pm 0.043$ & $2.04 \pm 0.11$  \\

$k T_{\rm e}$ (keV) 
& $3.106 \pm 0.056$ & $3.018 \pm 0.054$ & $3.5 \pm 1.1$
& $8.7^{+21}_{-1.9}$ & $20^{+61}_{-20}$ \\

$\tau$
& $10.89 \pm 0.30$ & $10.47 \pm 0.35$ & $6.9 \pm 3.0$ 
& $1.53 \pm 0.94$ & $0.43^{+3.0}_{-0.40}$ \\

$R_{\rm W}$ (km)
& $9.7 \pm 2.2$ & $ 9.8 \pm 2.1$ & $1.83 \pm 0.68$ 
& $2.77^{+0.53}_{-0.26}$ & $3.56^{+0.81}_{-0.40}$ \\

Photon Index 
& $1.83 \pm 0.37$ & $2.09 \pm 0.55$ & 2.09 (frozen) & 2.09 (frozen) 
& 2.09 (frozen) \\

Power-law N
& $0.015^{+0.051}_{-0.013}$ & $0.040^{+0.061}_{-0.010}$ 
& $< 0.027$  & $< 0.039$ & $< 0.018$ \\

$E_{\rm Fe}$ (keV)
& 6.6 (frozen) & $6.60 \pm 0.20$ & $6.65 \pm 0.11$
& $6.698 \pm 0.092$ & $6.58 \pm 0.16$ \\

$\sigma_{\rm Fe}$ (keV)
& $0.31^{+0.39}_{-0.07}$ & $0.65 \pm 0.33$ & $0.26 \pm 0.12$ 
& $0.21 \pm 0.17$ & $0.59 \pm 0.32$ \\

I$_{\rm Fe}$ $(\times 10^{-3})$
& $3.0 \pm 1.6$ & $4.9 \pm 3.6$ & $3.54 \pm 0.97$
& $3.1 \pm 1.0$ & $6.7 \pm 3.2$ \\

Fe Eq. W. (eV)
& 42 & 55 & 28 & 24 & 55 \\

$E_{\rm LE}$ (keV)
& $0.988 \pm 0.041$ & $1.064 \pm 0.045$ & $1.051 \pm 0.021$
& $1.045 \pm 0.013$ & $1.060 \pm 0.024$ \\ 

$\sigma_{\rm LE}$ (keV)
& $0.196 \pm 0.041$ & $0.184 \pm 0.060$ & $0.109 \pm 0.024$
& $0.124 \pm 0.020$ & $0.108 \pm 0.025$ \\

I$_{\rm LE}$ $(\times 10^{-2})$
& $9.4 \pm 2.0$ & $10.7 \pm 2.8$ & $5.41 \pm 0.96$ 
& $7.0 \pm 1.1$ & $7.7 \pm 1.8$ \\

LE Eq. W. (eV)
& 71 & 80 & 28 & 33 & 38 \\

Flux
& 1.17 & 1.40 & 1.97 & 2.08 & 1.94 \\

$\chi^2_{\rm red}$ (d.o.f.)
& 1.40 (193) & 1.24 (191) & 1.29 (190) & 1.68 (190) & 1.14 (189) \\

F-test
& $4.8 \times 10^{-4}$ & $3.6 \times 10^{-4}$ & 0.16 & 0.019 & 
0.78 \\

\hline
\end{tabular}
\end{center}
\end{table*}

The temperature of the disk blackbody component is $\sim 1.6$ keV, slightly
decreasing from the upper to the lower NB. The corresponding inner disk
radius, $R_{\rm in} \sqrt{\cos i}$, is between 8.1 and 9.7 km, slightly
increasing along the NB.  The Comptonized component gives a seed-photon 
temperature of $\sim 2$ keV, slightly decreasing along the NB. These photons
are Comptonized by hot electrons at a temperature that seems to increase from
$\sim 3.5$ keV to $\sim 20$ keV along the NB.  At the same time the optical 
depth of the Comptonizing cloud (calculated in the case of a spherical 
geometry) decreases from $\sim 7$ to $\sim 0.4$.  Note, however, that the  
uncertainty on these two parameters becomes very large in the lower-NB 
spectrum.  
We have calculated the radius, $R_{\rm W}$, of the seed-photon 
emitting region by using the bolometric luminosity of the soft photons 
obtained from the observed luminosity of the Comptonized spectrum after 
correction for energy gained in the inverse Compton scattering, and
assuming a spherical geometry.
This can be expressed as 
$R_{\rm W}=3 \times 10^4 D \sqrt \frac{f_{\rm bol}}{1+y}/
\left(kT_0 \right)^2$ km (in 't Zand et {\it al.} 1999), 
where $D$ is the distance to the source in
kpc, $f_{\rm bol}$ is the unabsorbed flux of the Comptonization spectrum in 
erg cm$^{-2}$ s$^{-1}$, $kT_0$ is the seed--photon temperature in keV, 
and $y=4kT_{\rm e}\tau^2/m_{\rm e} c^2$ is the relative energy gain due to the
Comptonization.  Inferred radii, reported in Table~1, 
are in the range between 1.8 and 3.5 km.  These are small when compared
to the values that were obtained for GX~17+2 (with $R_{\rm W} \sim 12-17$ km; 
Di Salvo et al. 2000) and GX~349+2 ($R_{\rm W} \sim 7-9$ km; Di Salvo et al. 
2001), by using a similar model.

We used the same continuum model to fit the spectra in the HB, although
in the spectrum of the upper HB the soft component is similarly well described 
by a blackbody or a {\tt diskbb} model. In fact, because the soft component 
has now a much lower temperature, these two models become virtually 
undistinguishable.
The inner disk temperature is now 0.8--0.9 keV, corresponding to a larger
inner disk radius of $R_{\rm in} \sqrt{\cos i} \sim 16-19$ km.
The seed-photon temperature is around 1 keV, with a radius of the
seed-photon emitting region of $R_{\rm W} \sim 10$ km. The Comptonizing cloud
has an electron temperature of $\sim 3$ keV and an optical depth of 
$\sim 10$.  The values of the parameters in this case are very similar to 
those found for GX~17+2 and GX~349+2.
  
\begin{figure*}
\vspace{0cm}
\hbox{\hspace{0cm}\psfig{figure=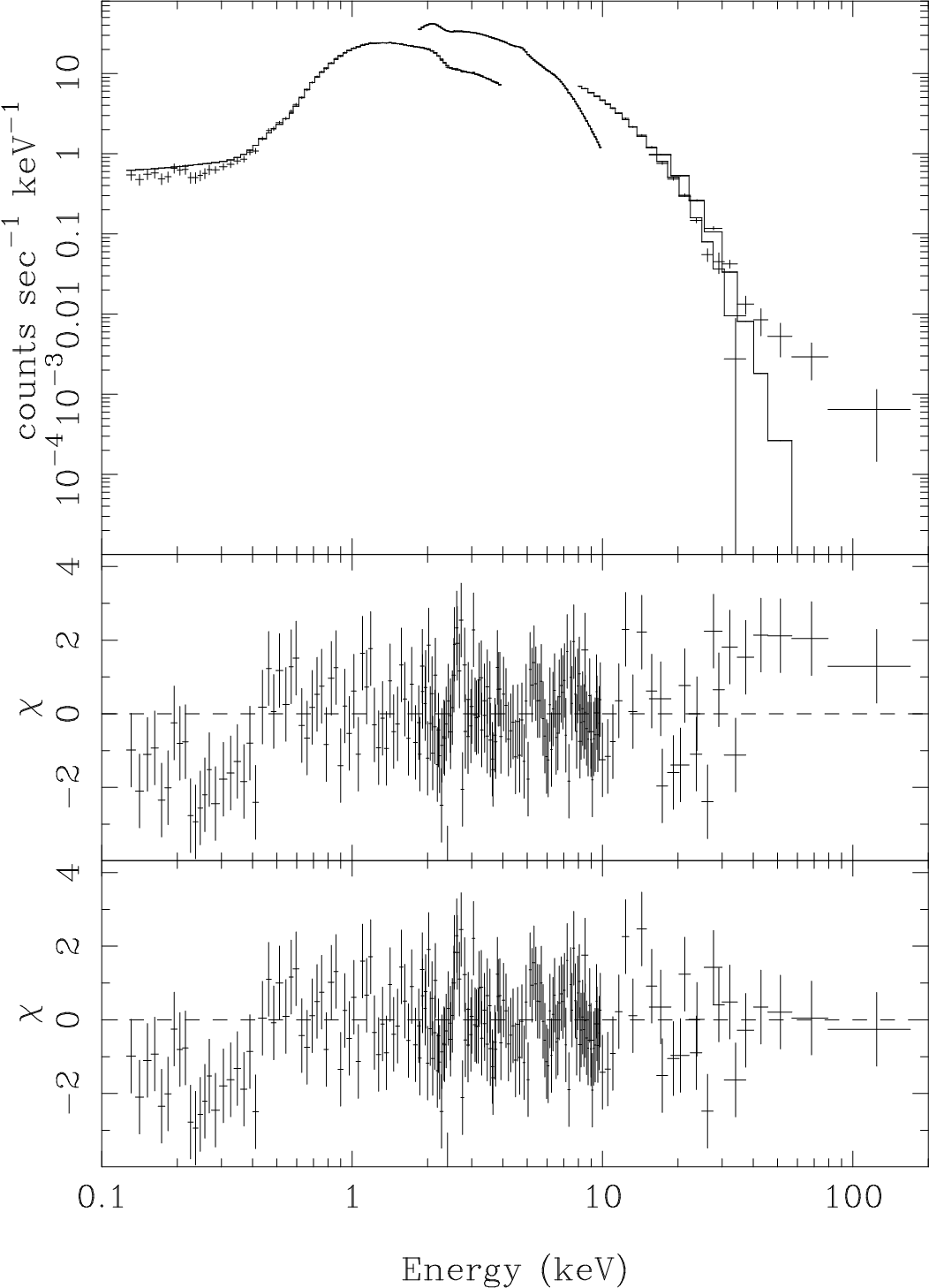,width=8.0cm}\hspace{1cm}
\psfig{figure=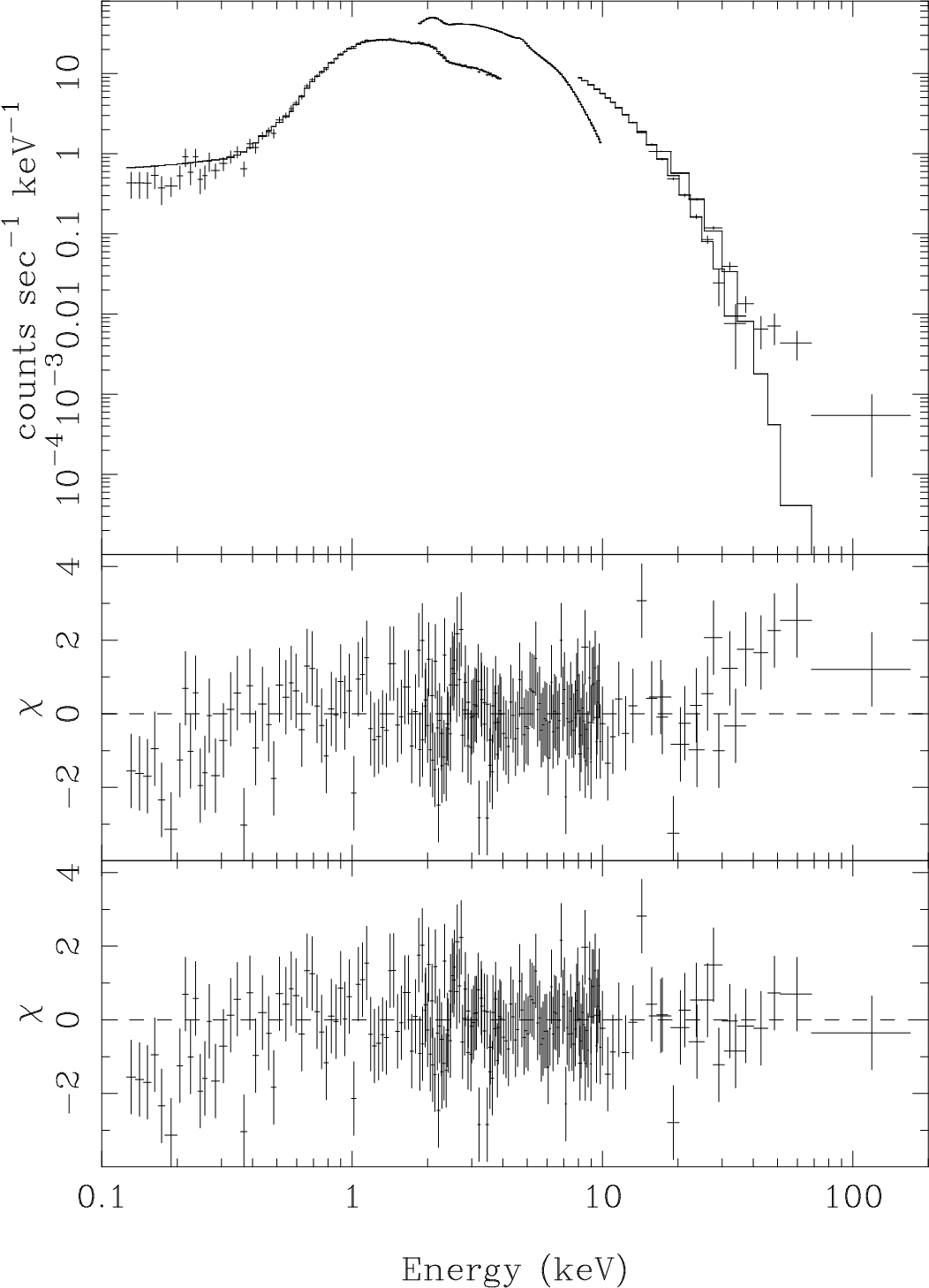,width=8.0cm}}
\vspace{0cm}
\hfill      \parbox[b]{18cm}{\caption[]{The broad band spectrum of Cyg X--2 
in the upper HB (left panel), and in the lower HB (right panel), respectively, 
together with the best fit two-component continuum model (disk blackbody plus 
{\tt comptt}) are shown in the top panels, and the corresponding residuals 
in unit of $\sigma$ are shown in the middle panels. 
Residuals in unit of $\sigma$ with respect to the best fit 
model reported in Table~1, including a hard power-law component, are
shown in the bottom panels.
}\label{fig3a}}%
\vspace{1cm}
\hbox{\hspace{0cm}\psfig{figure=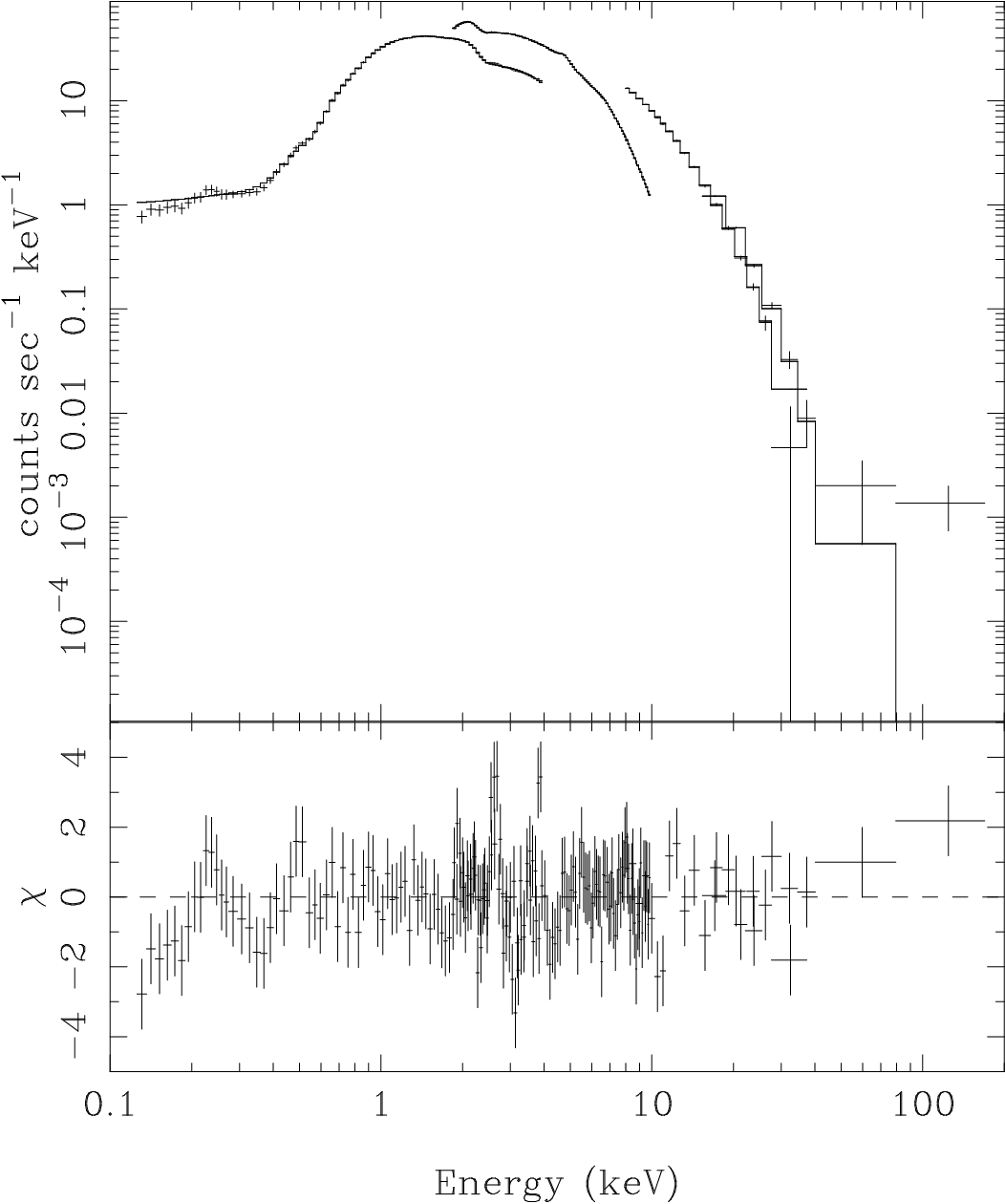,width=8.0cm}\hspace{1cm}
\psfig{figure=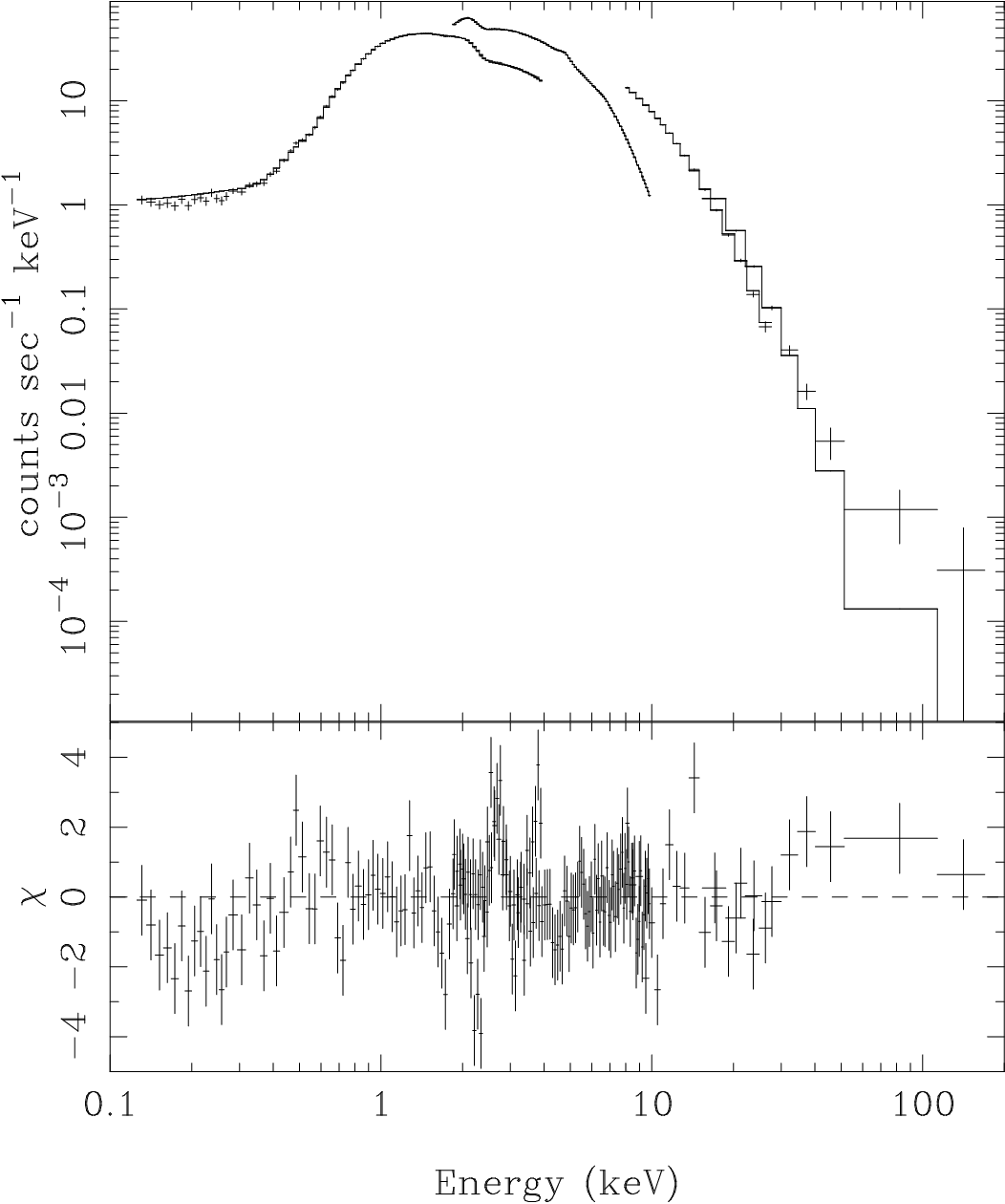,width=8.0cm}}
\vspace{0cm}
\hfill      \parbox[b]{18cm}{\caption[]{The broad band spectrum of Cyg X--2 
in the upper NB (left panel), and in the middle NB (right panel), respectively, 
together with the best fit model are shown in the top panels, and the 
corresponding residuals in unit of $\sigma$ are shown in the bottom panels. 
}\label{fig3b}}%
%
\end{figure*}

However, in these spectra the PDS data points above 30 keV were systematically 
above the model, as it can be seen in the residuals (in units of $\sigma$) 
shown in Fig.~\ref{fig3a} (middle panels). To fit these points
we added a power-law component to the continuum model. The addition of this
component eliminates the residuals at high energy (see Fig.~\ref{fig3a}, 
bottom panels), giving a reduction of the $\chi^2/d.o.f.$ from 292/195 to 
270/193 for the upper HB and from 259/193 to 234/191 for the lower HB,
corresponding to a probability of chance improvement (calculated using an 
F-test) of $\sim 4.8 \times 10^{-4}$ and $\sim 3.6 \times 10^{-4}$, 
respectively.
Other models for this hard excess (e.g., a thermal bremsstrahlung with 
temperature of the order of a few tens of keV) gave comparably good results.
The power-law component has a photon index of 1.8--2.1 and contributes 
$\sim 1.5\%$ of the 0.1--100 keV observed luminosity.  
We tried to average together the two spectra in the HB, given that
the obtained spectral parameters for these two spectra are quite similar. 
In this case, we could obtain a stable fit only fixing the $N_{\rm H}$ and
the Fe line centroid energy. The obtained spectral parameters for the 
average HB spectrum are similar to those obtained for each of the two 
spectra. However, in this case the statistical significance of the hard
power-law component is higher. The addition of this component in fact 
reduces the $\chi^2$ from 466 (263 d.o.f.) to 424 (261 d.o.f.), corresponding
to a probability of chance improvement of $\sim 4 \times 10^{-6}$.

Fixing the photon index
to 2.09 (i.e.\ the best fit value in the lower HB),
we found upper limits at 90\% confidence level to the 
power-law normalization in the spectra of the NB, where such a component is 
not required. These upper limits, together with the F-test for the addition 
of this component, are reported in Table 1.  
The derived upper limits are still 
compatible with the best fit values of the power-law normalization in the
HB, indicating that the reason for the non-detection of the power law in the
NB might be the hardening of the Comptonized component, the electron 
temperature of which systematically increases when the source moves from the 
HB to the NB.

Two emission lines are needed to fit all these spectra: the first one is at an
energy between 0.99 and 1.06 keV and seems to be stronger in the HB
(where it is also broader), with equivalent widths of 70--80 eV, than 
in the NB, where the equivalent widths are between 28 and 38 eV. 
The second emission line is at an energy of 6.6--6.7 keV, with equivalent 
widths from 24 to 55 eV. Note that also in this case the equivalent width is
higher when the width is larger (i.e.\ in the HB and lower NB, see Table~1).
Both these lines are highly significant in our spectra (F-tests give 
probabilities of chance improvement lower than $\sim 10^{-18}$ for the 
addition of the low-energy emission line, and lower than $7 \times 10^{-5}$  
for the addition of the iron K$\alpha$ line) and
were already known to be required to fit the spectrum of 
Cyg X--2 (see e.g.\ Kuulkers et al. 1997; Smale et al. 1993, and references 
therein).

\begin{figure*}
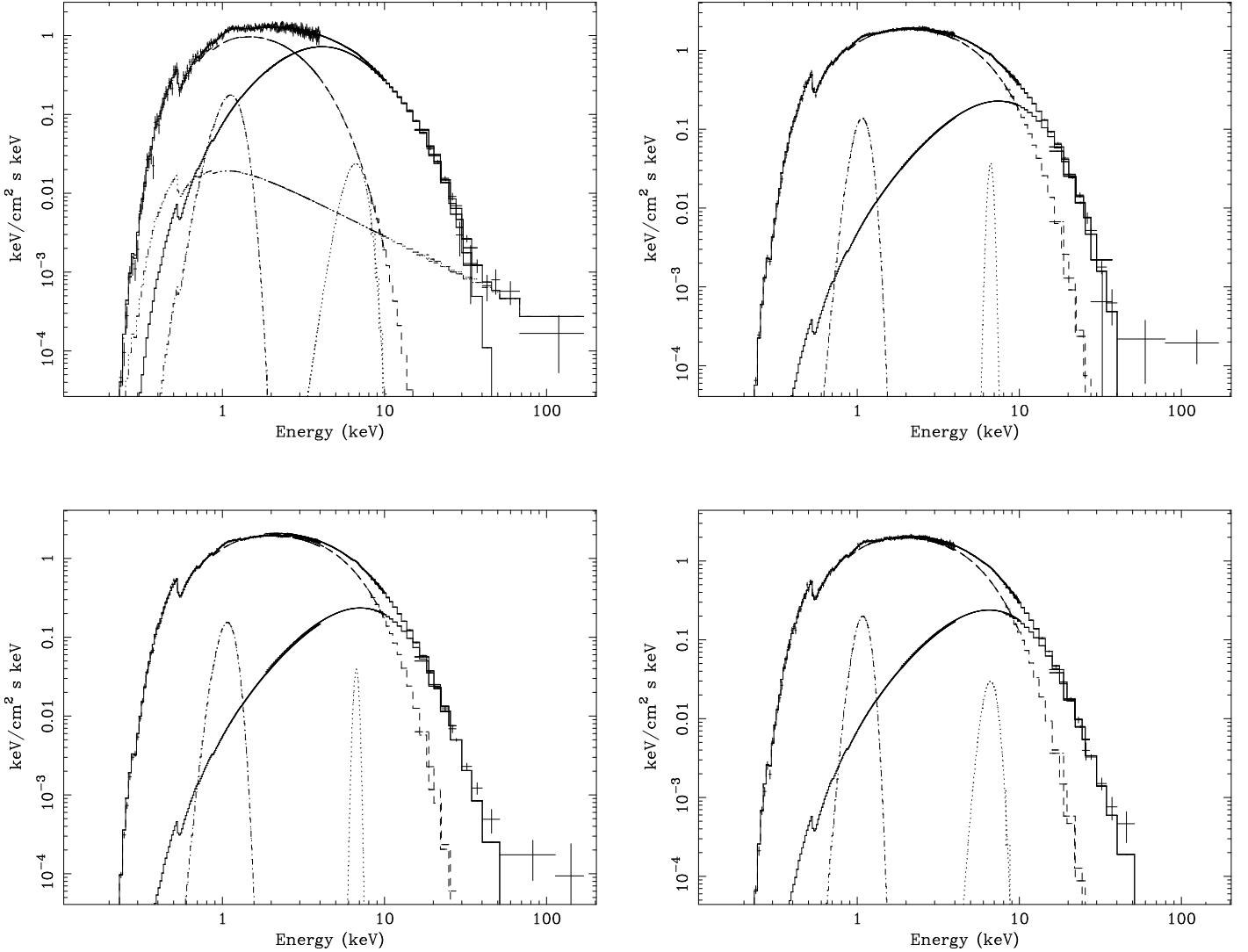

\vspace{0cm}
\hbox{\hspace{0cm}\psfig{figure=fig4a.ps,width=9.0cm,angle=270}\hspace{0.5cm}
\psfig{figure=fig4b.ps,width=9.0cm,angle=270}}
\vspace{1cm}
\hbox{\hspace{0cm}\psfig{figure=fig4c.ps,width=9.0cm,angle=270}\hspace{0.5cm}
\psfig{figure=fig4d.ps,width=9.0cm,angle=270}}
\vspace{0cm}
\hfill      \parbox[b]{18cm}{\caption[]{Unfolded spectra of Cyg X--2 in the lower HB (top left), upper NB
(top right), middle NB (bottom left), and lower NB (bottom right).
The best fit models reported in Table 1 are shown in this figure as the 
solid lines on top of the data. The individual model components are also 
shown, namely the disk blackbody (dashed line), the Comptonized spectrum 
({\tt comptt} model, solid line), two Gaussian emission lines, one at 
$\sim 1$ keV (dot-dashed line) and the other at $\sim 6.7$ keV (dotted line), 
and the power-law (when significantly detected, dot-dot-dot-dashed line).
}\label{fig4}}%
\end{figure*}

Data and residuals in units of $\sigma$ with respect to the corresponding
best fit model containing all the required spectral components 
are shown in Figs.~\ref{fig3a} and \ref{fig3b} (top and bottom panels, 
respectively) for four representative spectra.  Note that
for the upper-HB spectrum the fit is unstable if all parameters are set 
free, because of the interplay of the several spectral components to fit 
the low energy part of the spectrum. We obtained a stable fit by fixing the 
equivalent absorption column, $N_{\rm H}$, and the Fe line centroid energy 
to the best fit values obtained for the lower-HB spectrum.  
In all the other cases the fit was stable, including the lower-HB
spectrum which has the same number of parameters (this can be due
to the higher intensity in this interval).
To better see the spectral changes in Cyg X--2 according to the model
described above, we show in Fig.~\ref{fig4} the unfolded spectra for the 
interval corresponding to the lower HB and the three intervals 
in the NB, together with the components of the best fit model. 

From an inspection of the residuals (Fig.~\ref{fig3a} and \ref{fig3b}, bottom 
panels) we can see that some structures are still present at energies lower 
than 5 keV, which may indicate the presence of other narrow emission/absorption 
features.
In particular, the feature which has the highest statistical significance 
is an excess around 2.5 keV. Adding a Gaussian emission line at this energy
gives an improvement of the fit with a probability of chance improvement 
between $\sim 7 \times 10^{-2}$ and $\sim 10^{-6}$.
The highest statistical significance for the addition of this feature
is found in the middle NB spectrum, where this reduces the $\chi^2/d.o.f.$ 
from $320/190$ to $272/187$.
The best fit parameters of this line give a centroid energy of 
$2.631 \pm 0.033$ keV, a width $\sigma \la 0.09$ keV (compatible with the
instrumental energy resolution) and an intensity of 
$(7.3 \pm 1.8) \times 10^{-3}$ photons cm$^{-2}$ s$^{-1}$. 
Note that the reduced $\chi^2$ found for the middle NB spectrum is still
high to be statistically acceptable. However, the model  
described above is the one which gave the best fit to this spectrum and no 
obvious features are visible in the residuals with respect to this model.
Also a high value of the $\chi^2$ is not surprising given the high 
statistics of the BeppoSAX broad band spectra and the intrinsic complexity
of this kind of spectra.

\section{Discussion}

We have analyzed two BeppoSAX observations of the LMXB and Z source Cyg X--2
carried out in 1996 and 1997, respectively, for a total exposure time of
$\sim 100$ ks. As already mentioned, Cyg X--2 is known to show complex long 
term intensity variations; high-intensity and low-intensity states have been 
observed to be characterized by quite different behaviors. Cyg X--2 also shows 
secular shifts of the position of the Z-track in the CD/HID.  Therefore it 
is not straightforward to conclude whether or not the two BeppoSAX observations 
belong to similar states.  We have identified the BeppoSAX observation periods
in the RXTE/All Sky Monitor light curve.  We can see that the
1996 observation occurred a few days before the transition to a probable
low-intensity state, and probably corresponds to a ``medium state'',
which sometimes has been observed (Vrtilek et al. 1986).  The 1997 
observation occurred during a quite high-intensity state. However,
the CD and HID of these two observations connect smoothly
both in the hard and soft colors and the intensity, suggesting that 
there are no significant shifts between the two observations.  The source
was mainly in the HB during the 1996 observation and in the NB during the
1997 observation. We selected five spectra in different positions of the
CD/HID; two of these spectra correspond to the HB and three to the NB.
The 0.1--100 keV source luminosity increases from $0.89 \times 10^{38}$
ergs/s to $1.5 \times 10^{38}$ ergs/s (assuming a distance of 8 kpc) when 
the source moves from the upper HB to the lower HB and then to the upper NB. 
In the NB the luminosity remains almost constant at $1.5 \times 10^{38}$ ergs/s.

The X-ray spectrum of Cyg X--2 has a complex shape and requires several
components, which makes the spectral fitting non-unique (see also
Piraino et al. 2002; Done et al. 2002). Keeping this in
mind we now discuss the two-component continuum model which gave the best 
fit to most of the spectra analyzed here, which consists of a disk blackbody 
plus the Comptonization model {\tt comptt}. 
This is different from the continuum model that was
used to describe the BeppoSAX broad band spectrum of other Z sources
(GX~17+2, Di Salvo et al. 2000; GX~349+2, Di Salvo et al. 2001), i.e.\
a single temperature blackbody plus {\tt comptt}.  The difference might be
ascribed to the higher temperature of the soft component in Cyg X--2. 
In fact, while
in GX~17+2 and GX~349+2 the temperature of the blackbody component is in 
the range $0.5-0.6$ keV, in Cyg X--2 this temperature is higher, ranging from
$\sim 0.8$ keV in the upper HB up to $\sim 1.7$ keV in the NB. Note that when 
the temperature of the soft component is low (and close to the lower end of 
the instrument energy range) the disk multi-temperature blackbody and the 
single-temperature blackbody become indistinguishable, and in fact both
these models can well fit the Cyg X--2 spectrum in the upper HB, where the 
temperature of the soft component is the lowest.  On the other hand, when
the temperature of the soft component is high (and far from the lower end
of the energy range) the difference between the simple blackbody and the
disk blackbody (the low-energy end of which is considerably flatter) becomes 
important, and it is possible to discriminate between these models.
This leads us to interpret the soft component in all these spectra as 
the emission from the inner accretion disk, in agreement with the 
so-called Eastern model (Mitsuda et al. 1984).  
This interpretation is in agreement with the results obtained
from the spectral analysis of type-I X-ray bursts from GX~17+2
(Kuulkers et al. 2002).  In that case the soft component in the persistent 
emission remains constant during the bursts, thus suggesting that it does 
not come from the NS.  

Also the Cyg X--2 spectra 
show that the inner disk temperature significantly increases (from 
$\sim 0.8$ keV to $\sim 1.7$ keV) when the source moves from the HB to 
the NB, indicating that the inner rim of the disk approaches the NS at 
higher inferred mass accretion rates.  In fact the inner disk radius,
$R_{\rm in}$, obtained from these spectra decreases from 
$\sim 26$ km in the upper HB to $\sim 11.5$ km in the upper NB, assuming an
inclination angle $i = 60^\circ$.
Note that we are here neglecting the systematics effects affecting the
values of the inner disk inferred from the {\tt diskbb} model, such as the
color temperature correction (e.g.\ Shimura \& Takahara 1995), relativistic 
effects (e.g. Zhang et al. 1997) and the effect of the stress-free 
inner boundary condition (e.g.\ Frank et al. 1992), which usually 
tend to increase the inferred inner radii by a factor of the order of $1.5-2$.
However, the behavior of the inner disk radius
is at least qualitatively in agreement with the observed behavior of 
the frequencies of the kHz QPOs as a function of the position in the CD.
Two kHz QPOs were simultaneously observed in Cyg X--2, at $\sim 500$ and
$\sim 860$ Hz, respectively (Wijnands et al. 1998). The frequency of the
upper peak systematically increases when the source moves down the HB, 
reaching a maximum of $\sim 1$ kHz in the upper NB, before disappearing.
As a consistency check we can derive the NS mass from these independent
measurements.
Assuming that the lowest and highest frequencies measured for the upper
kHz QPO ($\sim 730$ Hz and $\sim 1000$ Hz, respectively) correspond, 
respectively, to the Keplerian frequency at the largest and smallest disk 
inner radius measured here, we can calculate the NS mass from the
Keplerian law: $M = \nu^2 R_{\rm in}^3 4 \pi^2/G$.
This gives $M \simeq (2.8 \pm 1.6) M_\odot$ and $M \simeq (0.45 \pm 0.03) 
M_\odot$, respectively, which are within a factor of 3 from the standard 
NS mass of $1.4 M_\odot$.
These are not unreasonable values considering the uncertainties in the 
disk model and in the correspondence between the kHz QPO frequencies and 
disk inner radii, which are not measured simultaneously in this case
(note that the errors on the inferred mass are derived considering only the 
uncertainties on the best fit values of the inner disk radius). In particular 
the radius in the NB is probably underestimated, but this is not surprising 
given that the correction factors mentioned above are expected to be higher 
when the radius becomes smaller.  Therefore the uncertainties are too large, 
at the moment, to allow us to prove or disprove the assumed model. 
  
This behavior, however, is different from that of GX~17+2 for which the 
parameters of the soft component seem to remain almost constant along the HB 
and NB (Di Salvo et al. 2000). This can also be deduced
from the shape of the CD of this source, where the HB and NB appear to be
almost vertical (with SC on the x-axis; see Wijnands et al. 1997), 
indicating that only small variations occur in the SC.  In other words,
in this scenario it is not clear why GX~17+2 does not show a variation 
of the soft component similar to the one observed for Cyg X--2, although 
it shows a similar behavior of the kHz QPO frequencies (see Wijnands
et al. 1997; Homan et al. 2002).  This different behavior of the inner
part of the disk might also have been observed in UV. Indeed UV observations 
of Sco X--1 (note that GX 17+2 is a Sco-like source, see below) show that the 
disk structure does not change when the source moves along the CD (Vrtilek 
et al. 1991). 
On the other hand UV observations of Cyg X--2 show that while the
outer disk structure remains unchanged, the inner disk (shorter UV 
wavelengths) varies when the source moves along the CD (S.D. Vrtilek 2001, 
private comm.).

Other significant changes at the transition between the HB and NB occur in the
parameters of the Comptonization component produced by the up-scattering of
soft seed photons by hot electrons. In particular, the seed-photon temperature 
increases from $\sim 1$ keV to $\sim 2.4$ keV when the source moves from the 
HB to the NB. At the same time the electron temperature of the Comptonizing 
plasma systematically increases 
and the optical depth systematically decreases. 
The radius of the seed-photon emitting region is $\sim 10$ km in the HB,
similar to the radii of the seed-photon emitting region inferred for
GX 17+2 and GX 349+2, while it decreases significantly in the NB, where
it ranges from $\sim 1.8$ km in the upper NB to $\sim 3.5$ km in the lower
NB.  The radius of the seed-photon emitting region suggests
that the NS itself (or the boundary layer between the accretion disk and the
NS) could provide the soft photons which are then Comptonized in a 
surrounding cloud. 
The Comptonizing cloud might then be a hot corona, a hot flared inner disk,
or even the boundary layer between the NS and the accretion disk if it is
sufficiently thin and hot (as suggested by Popham \& Sunyaev 2001).

However, in the NB of Cyg X--2 the seed-photon emitting radius becomes 
much smaller than the NS radius, although we note that the model we use
is probably too simple to realistically describe the emission region in
these systems, where gradients of temperature and other parameters are 
possible, if not probable.  The smaller value of the seed-photon radius 
is due to the bolometric flux of the Comptonized component, which is lower
by a factor of 2 with respect to that in the HB, and the seed-photon 
temperature, which is higher by a factor of 2. 

A possibility to explain the small seed-photon radius in the NB of
Cyg X--2 is that the geometry becomes non-spherical 
at some critical accretion rate. For instance, the fit results might be 
explained if at high accretion rates most of accreting plasma in the 
Comptonizing cloud collapses to a geometrically thin disk, leaving only an 
optically thin cloud around the inner part of the system, which can be 
responsible for the Comptonization. 
Because now the accretion occurs mostly onto the equator of the NS through 
the thin disk, the emitting region of soft seed-photons will be a strip
at the NS equator, implying a reduced emitting area (a geometry similar
to that proposed by Church \& Balucinska-Church 2001). Correspondingly,
because a similar (no more than a factor of two larger) amount of matter 
is accreted on a smaller area, the temperature of the NS blackbody emission, 
and therefore of the seed-photons for the Comptonization, 
increases.  However, it is not clear why the accreting plasma should collapse
to a geometrically thin disk 
at such high accretion rates.  Another possibility to explain 
these changes is that some occulting material (e.g. from a flared inner 
disk) is obscuring the inner part of the Comptonizing region.  In fact the 
absorption of the low energy part of the Comptonization spectrum might 
explain the reduction of the observed flux and  
the higher seed-photon temperature; the seed-photon temperature
can be overestimated in this case, because it is mainly determined by the 
low energy cutoff in the Comptonization spectrum. Note that also the values
we measure for the electron temperature and the optical depth may not be
representative of the average properties of the Comptonizing region if we 
mostly see the emission from the outer part, which could have a lower
optical depth and therefore a higher temperature. The fact that the Cyg-like
Z sources (i.e. Cyg X--2, GX 5--1 and GX 340+0) are thought to have a higher 
inclination than the other Z sources (referred to as Sco-like sources;
Kuulkers et al. 1994; Kuulkers \& van der Klis 1995) might explain why
in Cyg X--2 these changes are more pronounced than in GX 17+2. 

A hard tail, which can be fitted by a power law with photon index 
$\sim 1.8-2$, is significantly detected only in the HB spectra of 
Cyg X--2, 
and it is not required to fit the spectra in the NB.
This behavior is similar to that observed (with much higher statistical
significance, cf.\ Di Salvo et al. 2000) in GX 17+2, where a hard 
power-law tail was present in the HB spectra and weakened (by up to a
factor of $\sim 20$) when the source moved down the NB.
This similarity is in agreement with the idea of a correlation between
the radio flux, which is observed to have a maximum in the HB (Hasinger
et al. 1990; Penninx et al. 1988), and the hard tail, which 
would imply that the hard power-law component is related to the high-velocity 
electrons (probably from a jet) producing the radio emission. 
Note also that the power-law component seems to be weaker in Cyg X--2, 
where it contributes a lower fraction of the total luminosity, and, 
correspondingly, the radio emission is weaker (see e.g.\ Fender \& Hendry 
2000 and references therein).  This might be explained by the high 
inclination of the source, if the radio and the hard X-ray emissions are
produced by a jet which is collimated along the perpendicular to the disk.

Two (broad) emission lines are needed to fit the spectrum of Cyg X--2.
The first one is detected at an energy around 1 keV, with a
width of $\sigma \sim 0.1-0.2$ keV and equivalent width between 30 and 
80 eV.  Both the width and the equivalent width of this line are 
significantly larger in the HB than in the NB. 
The LECS data of the 1996 BeppoSAX observation, when the source was
in the HB, were already analyzed by Kuulkers et al. (1997).  They found
very similar parameters for this emission line, despite their somewhat 
different continuum model. This feature is probably due to blending of narrow
emission lines, as testified to by the fact that the previous higher-resolution
observations obtained with the Einstein Objective Grating Spectrometer
(OGS) require the presence of multiple line emission around 1 keV
from highly ionized Fe, O and other medium Z elements (Vrtilek et al. 1988).
The other emission line is detected at an energy of $6.6-6.7$ keV, compatible
with K$\alpha$ emission from highly ionized Fe. The width and equivalent
width of this line are also variable, in the range 0.2--0.65 keV and 
20--55 eV, respectively, without any clear correlation with the position
of the source in the CD/HID.  A broad (FWHM $\sim 1$ keV) Fe line at 
$\sim 6.7$ keV, with an equivalent width of $\sim 60$ eV, was measured in 
BBXRT data when the source was in a high state, on the lower part of the NB
(Smale et al. 1993).  These line parameters are very similar to 
the ones we measure in the lower NB spectrum. 
Smale et al. (1993) argue that this line can be produced by reflection from
the inner accretion disk if the disk surface is highly ionized.
Note that in this case the large width of the iron line is also in agreement 
with the idea that the source is observed at a high inclination angle.
Other narrow emission/absorption features could be present in the low
energy (below 5 keV) spectrum of Cyg X--2. In particular an emission
line at $\sim 2.5$ keV is detected with statistical significance. 
The energy of this line is consistent with Ly$\alpha$ emission from 
S~XVI at 2.62 keV (see also Kuulkers et al. 1997).

\section{Conclusions}

We have reported on the results of a broad band (0.1--200 keV) spectral
analysis of Cyg X--2 using two BeppoSAX observations taken in 1996 and 
1997, respectively.  The source was in the HB and NB during the 1996 and 
1997 observation, respectively. 
In our spectral deconvolution, using a two-component continuum consisting 
of a disk blackbody and a Comptonization model, the transition 
from the HB to the NB can be ascribed to a variation of the soft blackbody 
component, which becomes harder (and broader) in the NB, and of the 
Comptonization component, which becomes less prominent in the NB.  Based on 
these results, we interpret the soft blackbody component in the spectra of Z 
sources as the emission from the inner accretion disk, as previously proposed 
by the so-called Eastern model.  The Comptonization component may 
then be emitted by hot plasma surrounding the NS. In this interpretation, the 
changes in the parameters of the soft component indicate that the inner rim of 
the disk approaches the NS surface when the source moves from the HB to the NB, 
i.e.\ as the (inferred) mass accretion rate increases.  However, it is not 
clear why a similar behavior is not observed in other Z sources such as 
GX 17+2, where the soft component is not observed to change significantly as 
the source moves from the HB to the NB.
The parameters of the Comptonized component in Cyg X--2 also change 
significantly when the source moves from the HB to the NB. 
These changes can be explained by an occultation of the Comptonizing region 
probably caused by matter in an inner flared disk, or by an evolution from a 
spherical to a non-spherical geometry.
Finally we report the presence of a hard power-law tail, which is significantly 
detected in the HB spectra, where it contributes $\sim 1.5\%$ of the source 
luminosity. Although the relatively poor statistics does not allow a 
definitive conclusion, the hard tail might be weaker in the NB.  
This would be in agreement with the behavior shown by GX 17+2 and with the
idea of a correlation between radio flux and hard X-ray tails.

\begin{acknowledgements}
This work was partially supported by the Netherlands Organization for 
Scientific Research (NWO).
\end{acknowledgements}

\end{document}